\documentclass[prl,twocolumn,superscriptaddress]{revtex4}

\usepackage[utf8]{inputenc} 
\usepackage{graphicx} 
\usepackage{hyperref} 
\usepackage{xspace} 

\begin{document}

\title{On the interpretation of valence band photoemission spectra\\ at organic-metal interfaces}


\author{L. Giovanelli}
\affiliation{Aix-Marseille Universit\'{e}, CNRS, IM2NP UMR 7334, 13397, Marseille, France}

\author{F. C. Bocquet}
\affiliation{Aix-Marseille Universit\'{e}, CNRS, IM2NP UMR 7334, 13397, Marseille, France}
\affiliation{Peter Gr\"{u}nberg Institut (PGI-3), Forschungszentrum J\"{u}lich, 52425 J\"{u}lich, Germany}

\author{P. Amsalem}
\affiliation{Humboldt-Universits\"{a}t zu Berlin, Institut f\"{u}r Physik, D-12489 Berlin, Germany }

\author{H.-L. Lee}
\affiliation{School of Physical Sciences, Dublin City University, Ireland}
\affiliation{School of Chemical Sciences, Universiti Sains Malaysia, Penang, Malaysia}

\author{M. Abel}
\author{S. Clair}
\author{M. Koudia}
\author{T. Faury}
\affiliation{Aix-Marseille Universit\'{e}, CNRS, IM2NP UMR 7334, 13397, Marseille, France}

\author{L. Petaccia}
\author{D. Topwal}
\affiliation{Elettra Sincrotrone Trieste, Strada Statake 14 km 163.5, I-34149 Trieste, Italy}

\author{E. Salomon}
\author{T. Angot}
\affiliation{Aix-Marseille Universit\'{e}, CNRS, PIIM UMR 7345, 13397, Marseille, France}

\author{A. A. Cafolla}
\affiliation{School of Physical Sciences, Dublin City University, Ireland}

\author{N. Koch}
\affiliation{Humboldt-Universits\"{a}t zu Berlin, Institut f\"{u}r Physik, D-12489 Berlin, Germany }

\author{L. Porte}
\affiliation{Aix-Marseille Universit\'{e}, CNRS, IM2NP UMR 7334, 13397, Marseille, France}

\author{A. Goldoni}
\affiliation{Elettra Sincrotrone Trieste, Strada Statake 14 km 163.5, I-34149 Trieste, Italy}

\author{J.-M. Themlin}
\affiliation{Aix-Marseille Universit\'{e}, CNRS, IM2NP UMR 7334, 13397, Marseille, France}

\date{\today}

\begin{abstract}
Adsorption of organic molecules on well-oriented single crystal coinage metal surfaces fundamentally affects the energy distribution curve of ultra-violet photoelectron spectroscopy spectra. New features not present in the spectrum of the pristine metal can be assigned as "interface states" having some degree of molecule-substrate hybridization. Here it is shown that interface states having molecular orbital character can easily be identified at low binding energy as isolated features above the featureless substrate $sp$-plateau. On the other hand much care must be taken in assigning adsorbate-induced features when these lie within the $d$-band spectral region of the substrate. In fact, features often interpreted as characteristic of the molecule-substrate interaction may actually arise from substrate photoelectrons scattered by the adsorbates. This phenomenon is illustrated through a series of examples of noble-metal single-crystal  surfaces covered by monolayers of large $\pi$-conjugated organic molecules.
\end{abstract}

\maketitle

The nature of molecule-substrate interactions is of great relevance for the understanding of functional organic-inorganic hybrid systems\cite{Koch2008,Yang2003,Kakuta2007,Scholl2010,Ziroff2012,Puschnig2009}. The bonding at the interface can be studied by looking at the modification of the physical and chemical properties of the two constituents: the molecule and the substrate. Ultra-violet photoelectron spectroscopy (UPS) from the valence band is often used since it can map the density of states (DOS) of the sample surface through the energy distribution curve (EDC). The interface-induced DOS is then inferred by comparing the data of a thin film (often one monolayer (ML)) to those of the clean substrate and a thick molecular film, in which only weak intermolecular interactions play a role, thus providing the intrinsic molecular DOS. In this way the presence of molecular states belonging to molecular units in direct contact with the substrate can be identified. As an example, the filling of the lowest unoccupied molecular orbital (LUMO) level is detected as the appearance of new features at low binding energy in the EDC \cite{Koch2008,Scholl2010,Ziroff2012,Puschnig2011,Annese2007,Kilian2008,Kroger2010}. Such evidence is generally interpreted as charge transfer from the substrate to the molecules and can be corroborated by first principle calculations \cite{Ruiz2012}.

More generally, when the molecules are adsorbed on noble metal substrates, studying the binding energy (BE)  and line shape of frontier orbital levels is particularly useful in determining the molecule-substrate interaction \cite{Yamane2007,Puschnig2011}. Given their relatively low BE, the highest occupied molecular orbital (HOMO) levels often fall within the EDC of the rather featureless plateau of the single $sp$ electron band of the metal. By using angular-resolved photoemission this allows the angular distribution of photoelectrons to be probed, thus accessing properties such as the electron effective mass and interfacial orbital hybridization \cite{Yang2003,Yamane2007,Puschnig2009, Puschnig2011}.

The situation is more involved in the BE region of the metal $d$ band since possible molecular states are superposed to a very intense and structured metal-related EDC. Moreover, the scattering of substrate photoelectrons by the adsorbed layer can sensibly modify their spatial distribution and eventually their contribution to the spectrum \cite{Giovanelli2008,Giovanelli2010}. Despite this difficulty several studies reported on the formation of hybrid molecule-substrate interface states within the single crystal substrates $d$ band BE region\cite{Mariani2002, Evangelista2003, Baldacchini2004, Evangelista2004, DiFelice2004, Yamane2007, Baldacchini2006, Bussolotti2006, Baldacchini2007, DiCastro2002, Evangelista2009, Betti2010, Gargiani2010, Bussolotti2010,Bussolotti2010bis}. 

The present paper focuses on the substrate contribution in the EDC of the adsorbed system and it highlights the importance of the interface scattering when interpreting the EDC of a thin layer of large, $\pi$-conjugated organic molecules on noble-metal single-crystals substrates. As will be shown in the following, the adsorption promotes the measurements of EDCs in photoemission that are reminiscent of the metal 3D-integrated DOS, i.e., EDCs that mimic polycrystalline metal surfaces. At 1 ML coverage this contribution dominates the EDC of the clean substrate and should not be mistaken when assigning features to molecule-substrate interface states.

The approach adopted here consists in comparing UPS spectra of several adsorbed molecular systems with spectra measuring the substrate 3D DOS --namely the UPS of noble metal polycrystals-- thus revealing the striking similarities of the EDC of these two \emph{a priori} completely different types of systems.

\section{Experiment}
The data set presented is the result of various experimental runs performed on different ultra-high vacuum experimental systems. Experiments on silver surfaces were performed with synchrotron radiation at Elettra (BaDElPh beamline \cite{Petaccia2009}), BESSY (Suricat beamline) and Soleil (Antares beamline). When not otherwise stated, photoelectrons were collected in normal emission (NE) with linearly polarized light impinging at about 45 degrees from the sample surface normal. Samples were oriented with the high symmetry direction ($\Gamma$-X, $\Gamma$-K-X and $\Gamma$-L for (100), (110) and (111) surfaces respectively) in the scattering plane. Similar experimental geometries, but using unpolarized light from a He discharge lamp (He I line), were used for experiments performed on copper and gold. All spectra were recorded in angular integrated mode ($\pm 7^\circ$) with high resolution spectrometers. The BE scale is referenced to the substrate Fermi level. For data recorded with the He lamp, the contribution from satellite lines was subtracted from the spectra. Zinc-phthalocyanine (ZnPc), ZnPcF$_{8}$ and tetra(aminophenyl)porphyrin (TAPP) molecules were sublimated from quartz or tantalum crucibles onto single crystal substrates held at room temperature. The substrates were cleaned by several cycles of Ar ion sputtering and subsequent annealing. Molecular order was checked by low-energy electron diffraction (LEED) or scanning tunneling microscopy (STM).

\begin{figure}[!htp]  
\centering
\includegraphics[width=0.48\textwidth] {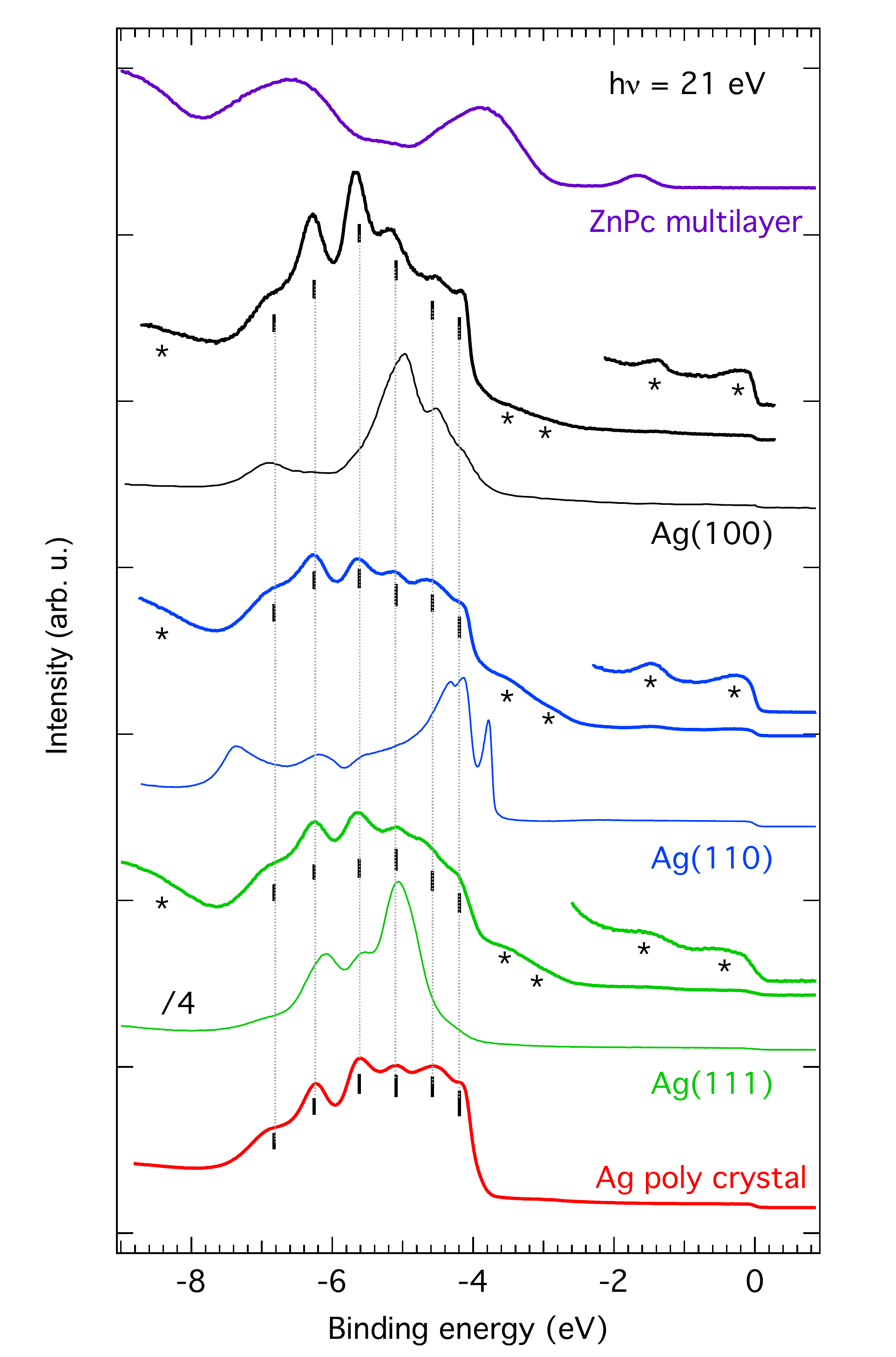} 
\caption{(color online). Normal emission, angle-integrated ($\pm 7^\circ$) photoemission spectra of 1 ML of ZnPc on Ag single-crystal surfaces. For every interface the clean sample spectrum is also displayed (thin curves). A blow up of the low binding energy region is shown to highlight the HOMO and LUMO contributions. For ZnPc/Ag(100) the low binding energy region is taken at $45^\circ$ emission angle. The spectrum of the clean Ag(100) is taken from Ref.\cite{Roloff1977}. The bottom and the top spectra correspond to a clean polycrystalline Ag sample and a multilayer of ZnPc, respectively. The spectra were normalized to their maximum intensity to ease the comparison.
\label{multiplot}	}	
\end{figure}

\section{Results}

The study of the adsorption of ZnPc on Ag(110) and (100) has been reported in detail \cite{Amsalem2009,Salomon2012}. The molecules adsorb with the macrocycle parallel to the surface adopting different geometrical structures depending on the coverage. In the present case the matrix transforming the substrate lattice vectors into the ones of the molecular superstructures are $\left(\begin{array}{cc}5 & 0 \\3 & 6\end{array}\right)$, $\left(\begin{array}{cc}4 & \pm2 \\2.5 & \mp3\end{array}\right)$ \cite{Amsalem2009}, and $\left(\begin{array}{cc}1 & 5 \\-5 & 1\end{array}\right)$ \cite{Salomon2012} for adsorption of ZnPc on Ag(111), (110) and (100) respectively.
Three sets of spectra corresponding to ordered MLs \cite{Amsalem2009, Salomon2012} of Zn-phthalocyanine (Zn-Pc) on the three low-index single crystal surfaces of silver are shown in Fig.\ref{multiplot}, together with the spectrum of a thick layer of ZnPc/Ag(110) (top) and that of an Ag polycrystalline sample (bottom). For reasons that will become evident in the following, vertical lines are drawn throughout Fig. \ref{multiplot} corresponding to the six main BE maxima features of the polycrystal spectrum. For each surface the clean substrate and the ZnPc ML spectra are shown. The spectra of the ZnPc multilayer grown on the three substrates (not shown) were virtually identical. 

As expected, the spectra of the three clean single crystal surfaces are markedly different from each other. Direct transitions as well as 1D DOS features contribute to a different extent to the observed EDC \cite{Roloff1977}. The spectra reproduce very well previously reported high-resolution results \cite{Roloff1977}. The silver polycrystal spectrum is representative of the Ag 3D DOS \cite{hufner}.

After the ZnPc ML adsorption the EDC of each sample is significantly modified. Interface molecular states (marked with an asterisk) are present at BE $< 4 eV$ superimposed on the flat \emph{plateau} of the substrate $sp$ states. These features are due to frontier orbitals such as (from higher to lower BE) the HOMO-2, HOMO-1, HOMO and partially filled LUMO \cite{Giovanelli2010}. The presence of a filled LUMO (sometimes referred to as former-LUMO) means that interface bonding involves electron charge-transfer from the substrate to the molecules. Filling of the LUMO is thus taken as a sign of strong interaction (chemisorption).
Other molecule-derived features are visible at BE higher than 7 eV, where the substrate contribution decreases. These are deep-lying molecular orbitals (MO) and in the present case do not participate in the molecule-substrate bonding. A shift to lower binding energy with respect to the thick film is observed for all MOs. This is due to the more effective final-state screening in the presence of the substrate metal electrons together with a redistribution of the MOs eigenenergies due to the filling of the LUMO.
 
EDC modifications are also very important within the BE range of the substrate 4$d$ band, that is from 4 eV to 7 eV BE. In fact in this region the three ML spectra are very similar, displaying six distinct features. Their BE and relative intensity coincide very well with that of the six features of the polycrystalline sample (vertical lines). This observation inhibits an assignment of the observed features as interface states and reveals that the substrate contribution to the measured EDC at the metal-molecule interface is that of a polycrystalline sample.

\begin{figure}[!htp]  
\centering
\includegraphics[width=0.48\textwidth] {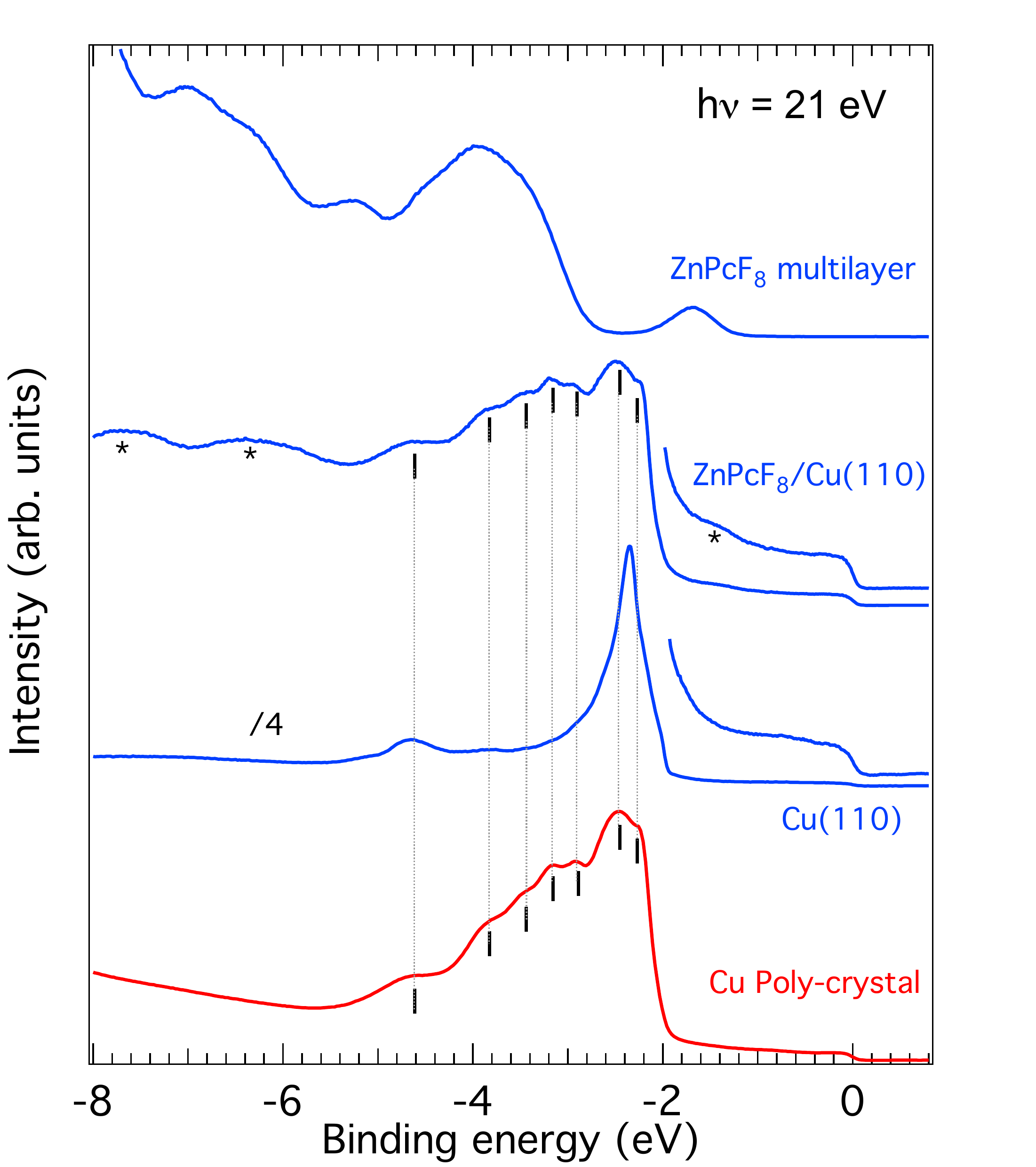} 
\caption{(color online). Normal emission, angle-integrated ($\pm 7^\circ$) photoemission spectra of ZnPcF$_{8}$ deposited on Cu(110). From bottom to top: a Cu polycrystalline sample, the clean Cu(110) surface, 1 ML ZnPcF$_{8}$/Cu(110), multilayer of ZnPcF$_{8}$/Cu(110).
\label{multiCu}	}	
\end{figure}

In Fig.\ref{multiCu} the spectra of a single disordered ML of ZnPcF$_{8}$/Cu(110) is displayed together with the clean substrate, the thick film and the Cu polycrystalline spectrum. Three clear molecular interface states are visible in the spectrum: the HOMO appearing as a weak feature at about 1.5 eV BE and two other deeper-lying MOs at 6.4 and 7.8 eV BE. No charge transfer to the LUMO is detected for this system. The intensity of the Cu(110) 4$d$ band is strongly quenched upon adsorption of ZnPcF8 and the EDC signature of the clean substrate is completely lost. Apart from a different background, (also due to the presence of spectral intensity from adsorbate MOs), the $d$ band region now looks just like that of the polycrystalline sample with all seven features (vertical bars) well reproduced in terms of BE and relative intensity.
As in the case of silver, such a close and striking resemblance indicates that the reported features are arising from a modified substrate contribution.

\begin{figure}[!htp]  
\centering
\includegraphics[width=0.48\textwidth] {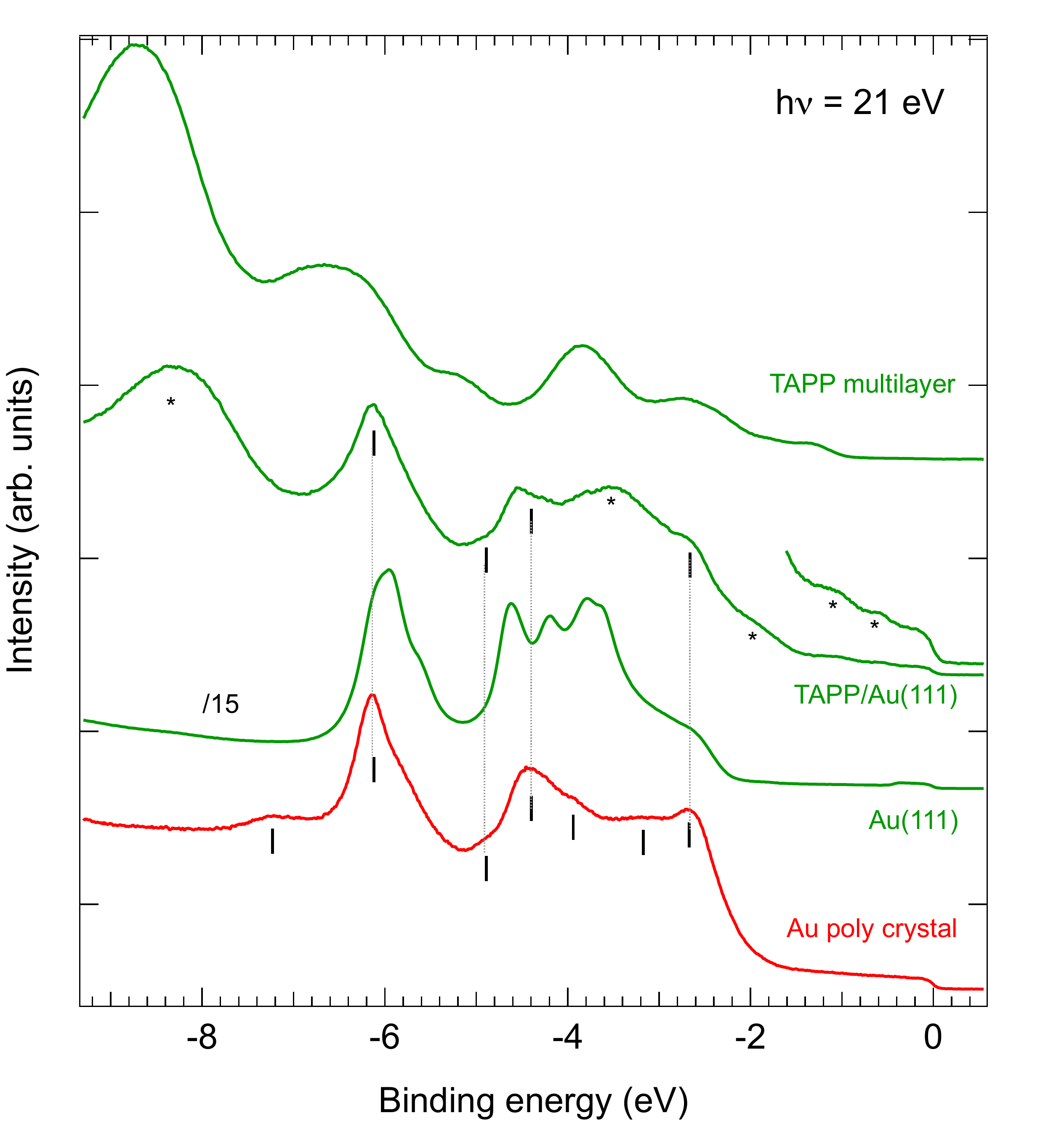} 
\caption{(color online). Normal emission, angle-integrated ($\pm 7^\circ$) photoemission spectra  of TAPP deposited on Au(111). From bottom to top: Au polycrystalline sample, clean Au(111) surface, 1 ML TAPP/Au(111), multilayer of TAPP/Au(111).
\label{multiAu}	}	
\end{figure}

Finally, an example of an organic molecule deposited on a gold single crystal is reported in Fig.\ref{multiAu}. The spectra of a single disordered layer of TAPP on Au(111), the multilayer, the bare substrate and the polycrystalline sample are compared. The relevant features of the latter are marked with vertical bars. From low to high binding energies they occur at 2.67, 3.15, 3.95, 4.4, 4.9, 6.14 and 7.2 eV respectively. At 1 ML coverage emission from several MOs is present in the spectrum: i) the low-BE features appearing on top of the substrate $sp$ plateau; ii) a feature at high BE (centered at 8.3 eV); iii) another one visible within the metal 5$d$ band at 3.5 eV. The latter is the intense feature appearing at 3.85 eV in the thick film spectrum that has shifted to lower BE in the ML spectrum. Once more the substrate contribution has changed significantly from that of the clean Au(111). In fact, the features between 3.4 and 4.4 eV are no longer discernible. Moreover, four of the substrate features coincide very well with those of a Au polycrystal spectrum (the other three being masked by the adsorbate contribution to the EDC). Consequently, they can not be assigned to interface states: they are the substrate contribution to the spectrum which is reminiscent of that of a polycrystal.

\section{Discussion}

The examples reported above are compelling evidence that adsorbing large $\pi$-conjugated organic molecules on noble-metal single crystals substantially changes the substrate contribution to the UPS spectrum. The latter should be considered as being close to that of a polycrystalline sample. 
Molecular levels are best detected at BEs where the substrate contributes the least, that is over the substrate $sp$ plateau and at BEs higher than the metal substrate $d$ band. Of course molecular features are also present within the BE region of the substrate $d$ band but their detection becomes more difficult because often they are masked by the overwhelming contribution of the substrate.

The fact that adsorbing a ML molecules transforms the EDC of the underlying single crystal to that of a polycrystalline sample can be understood as follows. Within the three-step model, photoemission in the UV occurs through direct transitions between bands in the crystal: the initial state is excited to a final unoccupied state lying at $h\nu$ eV higher energy, in the reduced zone scheme band structure. The reciprocal space vector through which the final-state band is back-folded guarantees momentum conservation. The electrons then travel to the surface and escape through the surface barrier before being detected by the analyzer in the vacuum. When atoms or molecules are adsorbed on a clean single crystal surface the escaping conditions of substrate photoelectrons are modified \cite{Anderson1976,Larsson1983,Lindgren1979,Bocquet2011,Giovanelli2010}. The overlayer --being ordered or disordered-- acts as a scattering layer which is able to provide parallel momentum and thus to change the emission directions of the photoelectrons. A photoelectron whose emission direction is off NE in the clean substrate can then be found at NE in the ML spectrum.  This phenomenon is sometimes referred to as surface-umklapp \cite{Anderson1976,Westphal1983,Larsson1983}. Equivalently, it can also be explained by the back folding of bulk (or surface) substrate bands by the reciprocal lattice vector of the adsorbed organic lattice\cite{Ziroff2009}.

When adsorbed on noble metal substrates, large $\pi$-conjugated organic molecules such as phthalocyanines or porphyrins normally orient with the molecular plane parallel to the surface. The typical lattice parameters are then about 1-2 nm. This corresponds to reciprocal lattice spacings of the order of 0.6-0.3 \AA$^{-1}$, which are very small when compared to the reciprocal lattice vector of the substrate. The parallel momentum transfer for photoelectrons with kinetic energies in the 10 - 15 eV range can then occur in several, closely-spaced directions. Eventually this results in a complete loss of angular anisotropy of the emitted photoelectrons as observed in Fig.\ref{multiplot} to \ref{multiAu}. Moreover, as demonstrated for the case of atomic adsorption, incoherent, quasi-elastic scattering can also average out the angular distribution of substrate photoelectrons \cite{Lindgren1979}. Spectra from noble-metal polycrystalline samples are, on the other hand, the sum of contributions from randomly-oriented micro-crystals. The lack of order in these samples results in a EDC reflecting the 3D DOS of the bulk crystal \cite{hufner}. The fact that the polycrystalline spectra and the spectra of single layers of phthalocyanine or porphyrins deposited on single-crystal surfaces resemble each other so closely reveals the effectiveness of the scattering (umklapp) process on the substrate photoelectrons at these interfaces. 

Interface scattering also affects the substrate $sp$ band emission \cite{Ziroff2009,Bocquet2011} of angle-integrated spectra. Whether this will result in distinct peaks or a modified background depends on photon energy and analyzer integration angle. In a recent low-photon energy (7-11 eV) ARPES study of ZnPc/Ag(110), it was shown that direct transitions in the $sp$-bands occurring away from the $\Gamma$ point are diffracted by the reciprocal lattice vectors giving rise to new, intense peaks at low BE in angle-integrated spectra measured around NE \cite{Bocquet2011}. At higher photon energy (e.g. 21 eV) the angular dispersion of the $sp$-bands will reduce \cite{Ziroff2009}. When using an integration angle of $\pm 7^\circ$, diffracted (or back folded) $sp$-bands are unlikely to show distinct peaks. Instead they may  modify the substrate background.

It is important to point out that interface scattering depends on the specific system under consideration through factors such as the adsorbate geometric order, the molecule-substrate interaction and the scattering power of the adsorbate \cite{Lindgren1979, Larsson1983}. When large molecules are adsorbed, the presence of several orientational domains is rather common. The present study shows that in this case, as in the case of disordered systems, the adsorbate promotes real 3D polycrystalline contributions from the substrate. When single domains are studied the reciprocal space lattice will become sparser and diffraction through selected reciprocal space rods can be observed. The presence of true polycrystalline features will depend on the effectiveness of incoherent scattering \cite{Lindgren1979}. 

The evidence reported above has consequences for the interpretation of the EDC modification following adsorption of organic molecules on single crystals.

Molecule-substrate interaction between $\pi$-conjugated molecules and noble-metal substrates often occurs through hybridization of frontier MOs with low-lying metal states  \cite{Koch2008, Kilian2008, Scholl2010, Giovanelli2008, Giovanelli2010}. Broadening of the HOMO and filling of the LUMO are often observed above the weak $sp$ band (see Fig. \ref{multiplot}-\ref{multiAu}). For high enough photon energies the $sp$ band generally contributes as a flat background in angle-integrated spectra. This has permitted the detailed study of the angular anisotropy of the molecular features giving insight into the details of molecule-substrate hybridization \cite{Puschnig2011}.

The interaction of molecular states with metal $d$ electrons is harder to unveil. This is mostly due to the fact that possible hybridization between molecular and metal states is blurred by the substrate contribution to the UPS spectrum. Still several UPS and ARPES studies have reported on the formation of interacting and hybrid states at the organic/noble metal interface $\emph{within the substrate d-band region}$ \cite{Mariani2002, Evangelista2003, Baldacchini2004, Evangelista2004, DiFelice2004, Yamane2007, Baldacchini2006, Bussolotti2006, Baldacchini2007, DiCastro2002, Evangelista2009, Betti2010, Gargiani2010, Bussolotti2010,Bussolotti2010bis}. In these studies intense and sharp photoemission features were found upon the adsorption of large (M-Pc, pentacene) and smaller (anthracene, small thiols, adenine) organic molecules. $Ab-initio$ calculations, polarization dependent and angular-resolved measurements were also reported generally highlighting the interface character of the photoemission features. Depending on the system considered, interface states were attributed to hybridization states or interaction states. 

In the reported studies when M-Pc or pentacene are adsorbed on Au(110)-(1x2) new features are observed within the Au 5$d$ bands \cite{Evangelista2003, Evangelista2004, Evangelista2009, Betti2010, Gargiani2010}. Because the features were not present neither for the clean substrate nor for the thick layer, they were assigned to hybrid interface states. Nevertheless, in angular integrated spectra these features have BE that are coincident with those of the Au polycrystal spectrum (see Fig.\ref{multiAu}) thus suggesting that their origin may arise from surface umklapp rather than molecule-metal hybridization.

When angular resolved experiments were performed on different M-Pc/Au(110)-(1x2), the dispersion of some of the features appearing at 1 ML coverage was assigned to mixed metal-molecule states and metal-mediated delocalization of molecular $\pi$-states \cite{Evangelista2009, Betti2010, Gargiani2010}. The fact that the dispersing features were coincident with that of polycrystalline Au was recognized but the diffraction of substrate photoelectrons was not addressed. It has to be noticed that angular dispersion from any "umklapped" state is  expected in the case of well-ordered single-domain molecular super-structures \cite{Bocquet2011,Ziroff2009}. Consequently, although angular dispersion from genuine molecule-metal mixed-states may take place in these systems, the modification of the EDC has to be benchmarked with a careful test on the role of interface diffraction from photoelectrons. This can be done, for instance, by measuring the ARPES spectra of the clean substrate at an angle corresponding to an overlayer reciprocal lattice vector (umklapp vector) \cite{Westphal1983}.

The presence of interface states within the metal $d$ bands was reported also for the adsorption on single crystal Cu surfaces of smaller molecules such as adenine \cite{Bussolotti2010}, anthracene \cite{Bussolotti2010bis} and 2-mercaptobenzoazole (MBO) \cite{Mariani2002, DiCastro2002}. For MBO/Cu(100) the experimental study was followed by a detailed theoretical investigation \cite{DiFelice2004, Ferretti2004}. If, on one hand, the bonding (dispersive) and anti-bonding (non-dispersive) states present at the sides of the Cu $d$ band were clearly identified, on the other, the origin of the features in between was less straightforward. Again, since their BE is coincident with that of the features of the Cu polycrystal spectrum (Fig.\ref{multiCu} of present paper), their origin could reside in the surface-umklapp process.

In the case of anthracene and adenine adsorption on Cu(110) hybrid interface states were recently reported \cite{Bussolotti2010, Bussolotti2010bis}. Among the features found at 1 ML coverage, those that were present neither in the clean Cu(110) nor in the thick film spectra were assigned to hybrid interface states. Interface hybridization lead to the conclusion that the molecules were chemisorbed. Actually, because their BE coincides with that of a polycrystalline spectrum (Fig.\ref{multiCu} of present paper), it is very likely that interface diffraction plays a major role in the promotion of the observed features. Consequently the hybrid character should be proved with another investigative technique such as, for example, resonant photoemission.

Finally, it should also be recalled that generally the appearance of a 3D DOS is related to the scattering power of the adsorbate and to the number of scattering rods through which the photoelectron can diffract. In cases such as 1D super-structures or for small molecules the number of diffracting rods accessible to substrate photoelectrons decreases. Eventually, the 3D DOS contribution will be taken over by well-defined peaks arising from direct transitions along a few rods of the molecular reciprocal lattice. This effect was studied in detail for the case of extra-emission in the $d$ band of noble-metal after adsorption of atomic species \cite{Westphal1983} but has not yet been addressed for the case of organic molecules.

\section{Conclusion}
By discussing several relevant examples it is proposed that interface scattering of substrate photoelectrons (surface-umklapp process or back folding of substrate bands) plays a crucial role in modifying the EDC when organic molecules are adsorbed on single crystal surfaces of noble metals. Particularly, in the case of large $\pi$-conjugated molecules the major effect is that the substrate contribution to the EDC becomes close to that of a polycrystalline sample. 
Diffraction of photoelectrons at organic-inorganic interfaces is expected to be always present and it has important implications for the interpretation of UPS spectra.

\section{Aknowledgements}
PA and NK acknowledge financial support by  the Helmholtz-Energie-Allianz "Hybridphotovoltaik" and the Sfb951 (DGF). AAC acknowledges financial support form Science Foundation Ireland (Grant number 09/IN.1/I2635)

\bibliography{refs}
\end{document}